%% file: guinn_lrt2015.tex
\documentclass[
    ,final            
  ]
  {aipproc}

\input{Majorana_author_LRT_2015}

\usepackage{multirow}

\layoutstyle{8x11single}

\begin{document}

\title{Low Background Signal Readout Electronics for the \textsc{Majorana Demonstrator}}

\classification{23.40-s, 23.40.Bw, 14.60.Pq, 27.50.+j}
\keywords      {neutrinoless double beta decay, germanium detector, low background electronics}

\mjauthorlist

\begin{abstract}
The \MJ\ Collaboration will seek neutrinoless double beta decay (\nonubb) in \iso{76}{Ge} using isotopically enriched p-type point contact (PPC) high purity Germanium (HPGe) detectors. A tonne-scale array of HPGe detectors would require background levels below 1 count/ROI-tonne-year in the 4~keV region of interest (ROI) around the 2039~keV Q-value of the decay. In order to demonstrate the feasibility of such an experiment, the \MJD, a 40~kg HPGe detector array, is being constructed with a background goal of $<3$~counts/ROI-tonne-year, which is expected to scale down to $<1$~count/ROI-tonne-year for a tonne-scale experiment. The signal readout electronics, which must be placed in close proximity to the detectors, present a challenge toward reaching this background goal. This talk will discuss the materials and design used to construct signal readout electronics with low enough backgrounds for the \MJD.
\end{abstract}

\maketitle


\section{Introduction to the \textsc{Majorana Demonstrator}}

The \MJ\ Collaboration will build an array of isotopically enriched, high purity Germanium (HPGe) detectors in order to search for neutrinoless double-beta (\nonubb) decay\cite{Avignone:2008} in \iso{76}{Ge}. Achieving a sensitivity to the \nonubb half-life of $>10^{27}$~years will require a tonne-scale array of detectors with a background rate below 1~count/ROI-t-y in the 4~keV region of interest around the 2039~keV Q-value for $\beta\beta$ decay in \iso{76}{Ge}. The \MJD\ is currently being constructed in order to demonstrate the feasibility of constructing such an array\cite{Abgrall:2014}. The \Demo\ will contain 40~kg of p-type point contact (PPC) HPGe detectors\cite{Barbeau:2007}, 30~kg of which will be composed of Germanium 87\% isotopically enriched in \iso{76}{Ge}. Two separate modular arrays consisting of 7 strings of 4 or 5 detectors will be constructed in order to demonstrate scalability to a tonne-scale array. The \MJD\ has a background goal of $<$3~counts/ROI-t-yr, which is expected to scale down to $<$1~count/ROI-t-y due to effects such as self-shielding, based on simulations.

\begin{figure}[t] \label{MJD}
  \centering
  \includegraphics[width=.7\textwidth]{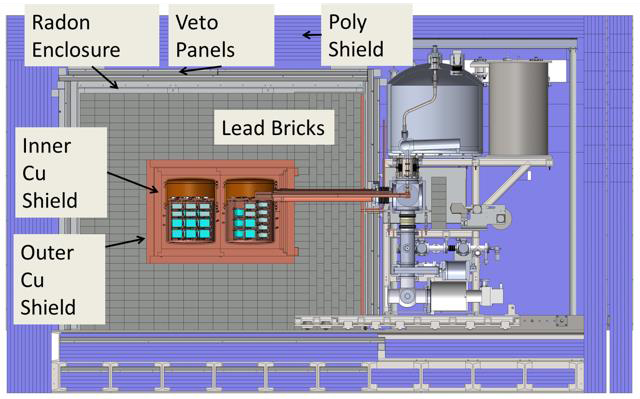}
  \caption{The configuration of the \MJD, with both cryostats placed inside of the shielding.}
\end{figure}

In order to achieve such a low background rate, multiple approaches are taken. The experiment will take place at the 4850~ft level of Sanford Underground Research Facility (SURF), in Lead, South Dakota, providing a 4260~mwe overburden to shield from cosmic rays. The detector modules will be housed in passive shielding against environmental background sources consisting of high density polyethylene, lead, and copper. Active background rejection approaches include a muon-veto system and rejection of multi-site events. Most $\gamma$-ray backgrounds will appear as multi-site events due to Compton scattering, and the low electric fields provided by PPC detectors produce multiple rises in these events that can be found by various analysis techniques. Furthermore, all components inside of the shielding will consist of radio-pure materials. In particular, the copper that forms the structure of the array will be ultra-pure electroformed copper (EFCu), produced and machined on the 4850~ft level of SURF. The amount of material placed inside of the shielding other than the HPGe detectors and EFCu is minimized.

In order to verify the purity of materials used for the \MJD, an extensive radio-assay and background-modeling campaign has been carried out. Four assay-techniques were used to determine the purity of materials:

\begin{itemize}
\item Direct $\gamma$-counting uses HPGe detectors to measure a background spectrum from components. This took place at underground facilities at the Waste Isolation Pilot Plant (WIPP) near Carlsbad, NM, the Kimballton Underground Research Facility (KURF) near Ripplemead, VA, and the low background facility (LBF) at Oroville, CA. The Oroville facility has recently moved to SURF.
\item Inductively coupled plasma mass spectrometry (ICP-MS) was performed at facilities at Pacific Northwest National Laboratory (PNNL) and Lawrence Berkeley National Laboratory (LBNL). ICP-MS is the most sensitive technique to uranium and thorium backgrounds used by the \MJD.
\item Glow discharge mass spectrometry (GDMS) was performed by the National Research Council of Canada (NRCC). GDMS offers high sensitivity to uranium and thorium content in conductive materials and was used primarily for the lead shielding.
\item Neutron activation analysis (NAA) was performed, with test samples neutron activated at research reactors at the University of California, Davis' McClellan Nuclear Radiation Center, North Carolina State University and Oak Ridge national Laboratory. Activation product levels were then measured at each of these sites or by the previously mentioned $\gamma$-counting facilities. NAA is used for low atomic mass materials such as plastics.
\end{itemize}

\begin{figure}[t] \label{MJDBkgs}
  \centering
  \includegraphics[width=.7\textwidth]{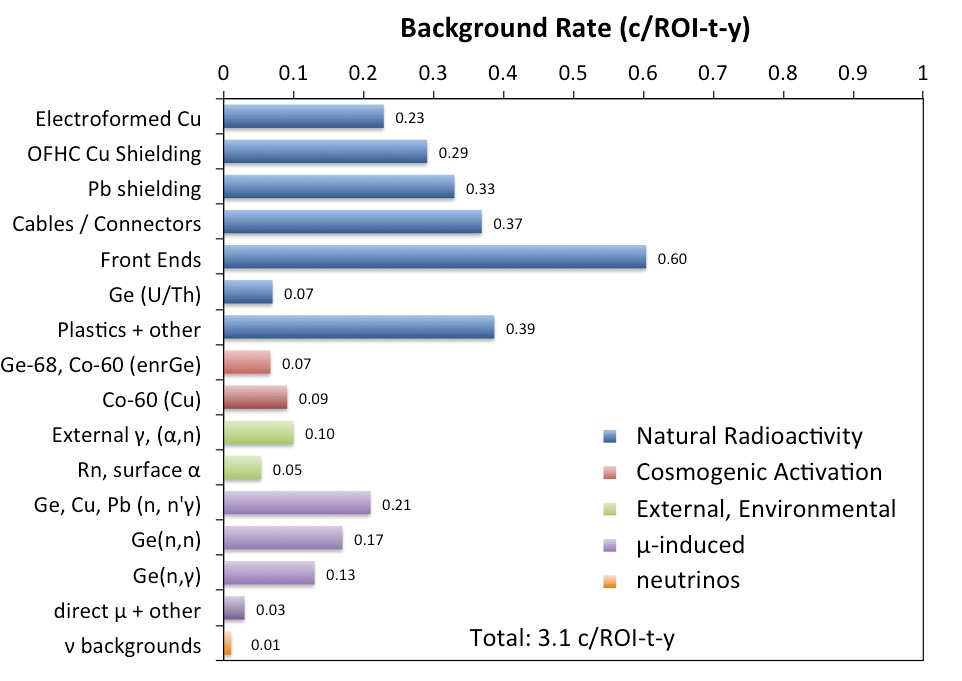}
  \caption{A summary of predicted background for the \MJD, produced in September 2014. The net backgrounds are 3.1 counts/ROI-t-yr, with some of the largest contributions coming from signal readout electronics.}
\end{figure}

The detection efficiency of each detector to backgrounds from each component of the \MJD\ is estimated using a detailed simulation of the detector and shield geometry performed using Geant4\cite{Cuesta:2014}. The results of this simulation are combined with the assay results to obtain an estimate of the total background levels of the detector. In March, 2015, the expected background level was $<$3.1~counts/ROI-t-yr, with individual contributions from various components listed in Figure \ref{MJDBkgs}. The largest background contributions come from the signal readout electronics, consisting of the front-end boards, signal cables and signal connectors.

\section{Signal Readout Electronics}
\begin{figure} \label{Cable}
   \hfil \includegraphics[width=.4\textwidth]{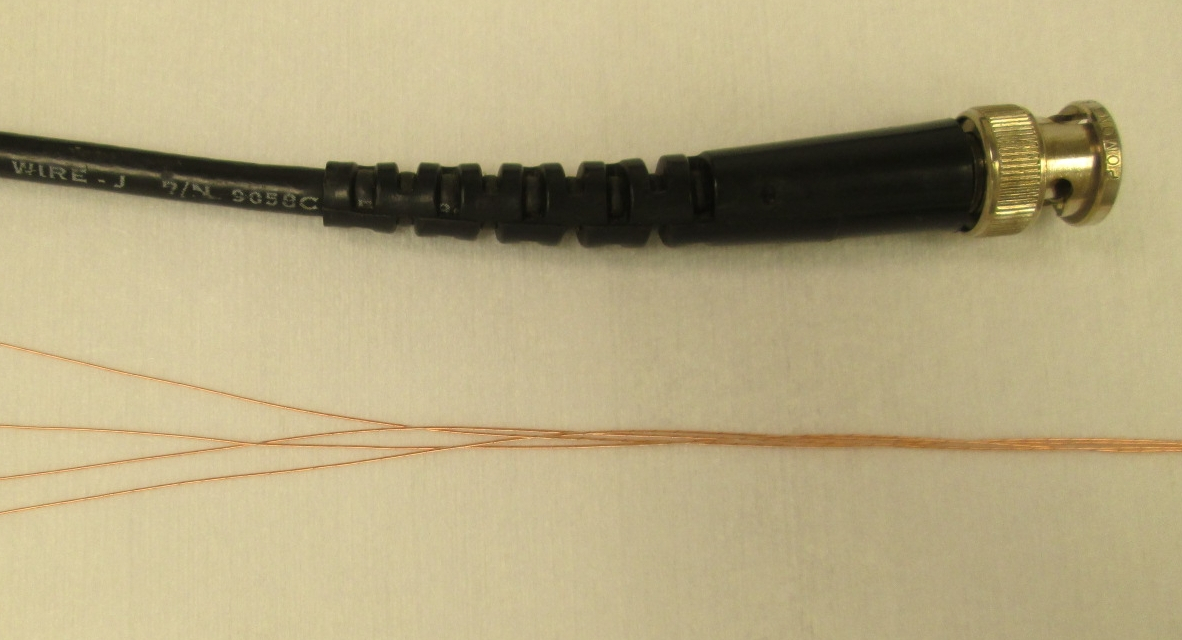}
   \hfil \includegraphics[width=.33\textwidth]{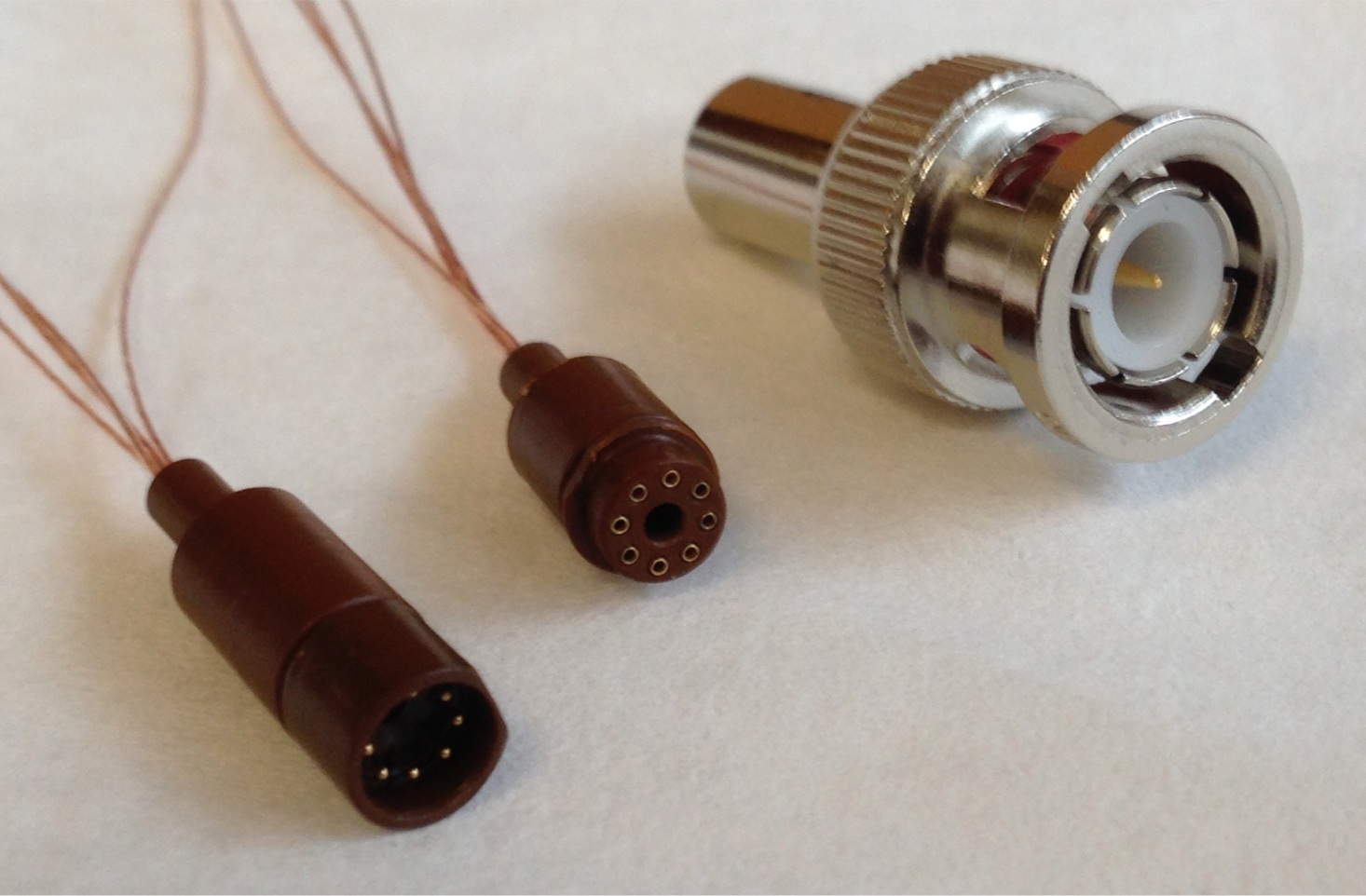}
   \hfil \includegraphics[width=.23\textwidth]{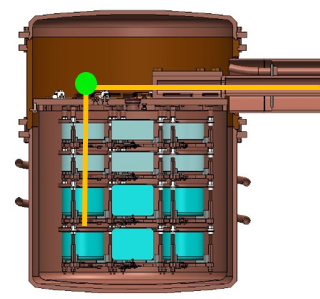}
   \caption{Left: Bundle of four signal cables, with a 0.195~in coaxial cable for comparison. Middle: Signal plug pair, with a BNC connector for comparison. Right: the path taken by the cables through the cryostat. The dot represents a signal connector.}
\end{figure}

The signal readout electronics chain is responsible for collecting the charge from the Germanium detectors and carrying the signal to preamplifiers and digitizers outside of the shielding. The charge is collected by a contact pin at the point contact and shaped by front-end boards, which sit directly adjacent to the point contact of the detectors and act as the first stage of a charge sensitive amplifier for the signal. A bundle of four coaxial cables, called a detector bundle, carries the signal up along a string to the cryostat cold plate. A second bundle, the feedthrough bundle, carries the signal from the cold plate, along the thermosyphon crossarm, to a feedthrough flange outside of the shielding. The two cable bundles are connected by signal connectors directly above the cold plate. The path taken by the cables is depicted in Figure \ref{Cable}.  Outside of the feedthrough flange, the signal is fed into preamplifiers and then digitized. Because the front-end boards, cable bundles, and signal connectors are all located inside of the shielding, it is critical that all three use a low-mass design with radiopure components in order to meet the \Demo's background goals. The components must also be robust under vacuum cycling and thermal cycling to liquid nitrogen temperatures, and must be easy to use in a glove box without breaking.

\subsection{Low-Mass Front-End Boards}
\begin{figure} \label{LMFE}
   \hfil \includegraphics[width=.45\textwidth]{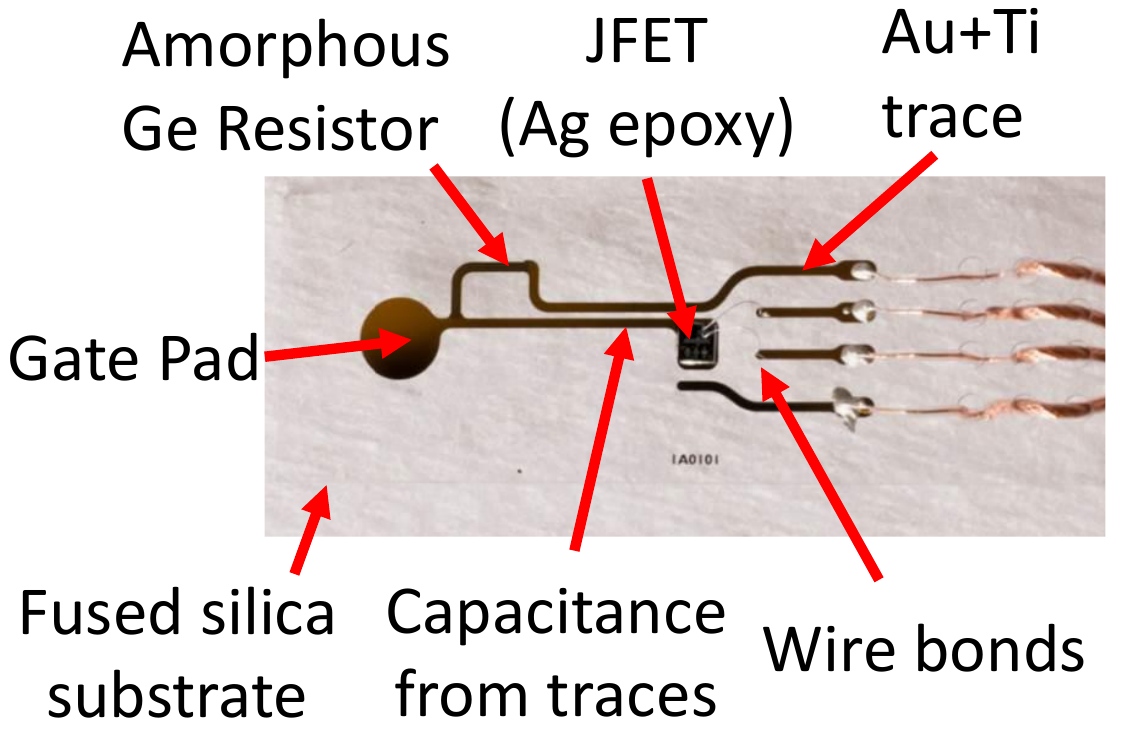}
   \hfil \includegraphics[width=.45\textwidth]{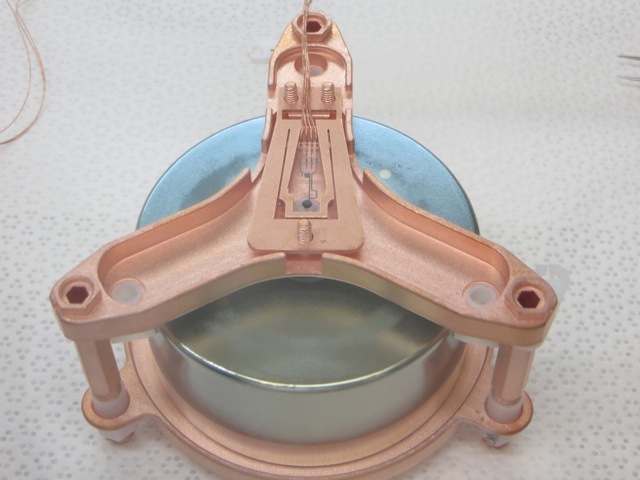}
   \caption{Left: LMFE board with each part labeled. Right: LMFE board mounted on a detector assembly.}
\end{figure}

The front-end boards use a design developed for the \MJD, called low-mass front-ends (LMFEs). The LMFEs are charge-sensitive amplifiers that use a resistive feedback design \cite{Barton:2011}. A charge-injection line is provided to allow characterization and monitoring of the LMFEs by external pulsers. The circuit is fabricated on a fused silica substrate. Gold traces are printed on the substrate on a titanium adhesion layer using photolithography. The feedback and charge-injection capacitances are provided by the geometric configuration of the traces. The feedback resistor consists of amorphous germanium, which can be made more radiopure than more common ceramic resistors using the same techniques used to produce HPGe detectors. A bare JFET die is attached to the substrate using low background silver epoxy, and connected to the circuit traces via wire bonding. The same Ag epoxy is used to attach the cable bundles to the LMFE. The LMFE board is mounted on an EFCu spring clip, which can be tensioned to provide a contact force between the gate pad of the LMFE circuit, a contact  pin, and the point contact of a germanium detector. Figure \ref{LMFE} contains two photographs of an LMFE. ICP-MS assay of a full LMFE and spring clip has been performed, finding activities from uranium and thorium of 1.4~$\mu$Bq/LMFE, resulting in an expected 0.60~counts/ROI-t-yr. Assays have also been performed on each individual material, shown in Table \ref{LMFEBkgs}, resulting in an expected activity of $<$~715~nBq/LMFE or $<$0.27~counts/ROI-t-yr. This discrepancy is most likely due to contamination from the handling of the materials during LMFE production. Increased care has been taken during subsequent LMFE production that may reduce backgrounds further.

\begin{table}[htbp] \label{LMFEBkgs}
   \centering
   \input{LMFEBGTable}
   \caption{Purity and background contribution of each material component of LMFEs.}
\end{table} 

\subsection{Signal Cables}
The electrical signal is carried from the LMFE to a feedthrough flange through a 1.5~m bundle of four coaxial cables. The cables have an outer diameter of only 0.4~mm, and a linear mass density of 0.4~g/m. The signal cable is custom made by Axon' Cable S.A.S. in France using commercially available clean copper. EFCu is not necessary due to the small size of the cables. The cables have an impedance of 50~$\Omega$ for up to 2.2~m. Radio-assay of a short length of cable found activity of 0.059~$\mu$Bq/m, contributing an estimated 0.085~counts/ROI-t-yr.

\subsection{Signal Connectors}
Signal connectors are necessary to connect the detector cable bundles to the feedthrough cable bundles above the cold plate of the cryostat. Commercially available electrical connectors cannot attain low enough backgrounds because they contain beryllium-copper (BeCu) electrical contact springs, which is difficult to purify of uranium and thorium to levels acceptable by the \MJD. Instead, a novel plug design with contact pins that contain no contact springs was used. The pins and sockets are misaligned so that the pin is forced to bend as it is inserted, and the spring action of the pin provides contact force. Mill-Max\textsuperscript{\textregistered} gold-plated brass pins were used, with the contact springs removed. The pins are held in a Vespel\textsuperscript{\textregistered} housing. The cables are attached using a low background tin-silver eutectic developed for the SNO experiment \cite{Amsbaugh:2007}. Strain relief is provided by fluorinated ethylene propylene (FEP) shrink tubing. Extensive quality control testing is necessary in order to ensure reliable connections. Full body assay of two signal connectors found activities of 1525~$\mu$Bq/connector-pair, leading to a predicted 0.284~Counts/ROI-t-yr. Individual material assays, shown in Table \ref{PlugBkgs} have found activities of $<$2110~$\mu$Bq/connector-pair, with the largest component coming from the pins and sockets. Notice that the addition of contact springs would single-handedly triple \MJD's background goal.

\begin{table}[htbp] \label{PlugBkgs}
   \centering
   \input{PlugBGTable}
   \caption{Purity and background contribution of each material component of the signal connectors.}
\end{table} 

\section{Conclusion}
In order to reach the \MJD's background goal of $<$3~counts/ROI-t-yr, novel designs for signal readout electronics are necessary. Through miniaturization and the use of low-radioactivity materials, the low-mass front-end boards, signal bundles and BeCu-free signal connectors were created, helping to reduce the total background estimation to $<$3.1~counts/ROI-t-yr.

\begin{theacknowledgments}
This material is based upon work supported by the U.S. Department of Energy, Office of Science, Office of Nuclear Physics. We acknowledge support from the Particle Astrophysics Program of the National Science Foundation. This research uses these US DOE Office of Science User Facilities: the National Energy Research Scientific Computing Center and the Oak Ridge Leadership Computing Facility. We acknowledge support from the Russian Foundation for Basic Research. We thank our hosts and colleagues at the Sanford Underground Research Facility for their support.
\end{theacknowledgments}



\bibliographystyle{aipproc}   

\end{document}

%% file: Majorana_author_LRT_2015.tex
\newcommand{\MJ}{M{\sc ajo\-ra\-na}}
\newcommand{\Demo}{D{\sc e\-mon\-strat\-or}}
\newcommand{\MJD}{{\MJ\ \Demo}}
\newcommand{\nonubb}{0$\nu\beta\beta$}
\newcommand{\iso}[2]{$^{#1}$#2}

\newcommand{\blhill}{Department of Physics, Black Hills State University, Spearfish, SD, USA}
\newcommand{\ITEP}{Institute for Theoretical and Experimental Physics, Moscow, Russia}
\newcommand{\JINR}{Joint Institute for Nuclear Research, Dubna, Russia}
\newcommand{\lbnl}{Nuclear Science Division, Lawrence Berkeley National Laboratory, Berkeley, CA, USA}
\newcommand{\lanl}{Los Alamos National Laboratory, Los Alamos, NM, USA}
\newcommand{\uw}{Center for Experimental Nuclear Physics and Astrophysics, 
and Department of Physics, University of Washington, Seattle, WA, USA}
\newcommand{\unc}{Department of Physics and Astronomy, University of North Carolina, Chapel Hill, NC, USA}
\newcommand{\duke}{Department of Physics, Duke University, Durham, NC, USA}
\newcommand{\ornl}{Oak Ridge National Laboratory, Oak Ridge, TN, USA}
\newcommand{\ou}{Research Center for Nuclear Physics and Department of Physics, Osaka University, Ibaraki, Osaka, Japan}
\newcommand{\pnnl}{Pacific Northwest National Laboratory, Richland, WA, USA}
\newcommand{\ttu}{Tennessee Tech University, Cookeville, TN, USA}
\newcommand{\sdsmt}{South Dakota School of Mines and Technology, Rapid City, SD, USA}
\newcommand{\usc}{Department of Physics and Astronomy, University of South Carolina, Columbia, SC, USA}
\newcommand{\usd}{Department of Physics, University of South Dakota, Vermillion, SD, USA} 
\newcommand{\ut}{Department of Physics and Astronomy, University of Tennessee, Knoxville, TN, USA}
\newcommand{\tunl}{Triangle Universities Nuclear Laboratory, Durham, NC, USA}

\newcommand{\mjauthorlist}{
\author{I.~Guinn$^{a}$, N.~Abgrall$^{b}$, I.J.~Arnquist$^{c}$, F.T.~Avignone~III$^{d,e}$, C.X.~Baldenegro-Barrera$^{e}$, A.S.~Barabash$^{f}$, F.E.~Bertrand$^{e}$, A.W.~Bradley$^{b}$, V.~Brudanin$^{g}$, M.~Busch$^{h,i}$, M.~Buuck$^{a}$, D.~Byram$^{j}$, A.S.~Caldwell$^{k}$, Y-D.~Chan$^{b}$, C.D.~Christofferson$^{k}$, C. Cuesta$^{a}$, J.A.~Detwiler$^{a}$, Yu.~Efremenko$^{l}$, H.~Ejiri$^{m}$, S.R.~Elliott$^{o}$, A.~Galindo-Uribarri$^{e}$, T.~Gilliss$^{n,i}$, G.K.~Giovanetti$^{n,i}$, J. Goett$^{o}$, M.P.~Green$^{e}$, J.~ Gruszko$^{a}$, V.E.~Guiseppe$^{d}$, R.~Henning$^{n,i}$, E.W.~Hoppe$^{c}$, S.~Howard$^{k}$, M.A.~Howe$^{n,i}$, B.R.~Jasinski$^{j}$, K.J.~Keeter$^{p}$, M.F.~Kidd$^{q}$, S.I.~Konovalov$^{f}$, R.T.~Kouzes$^{c}$, B.D.~LaFerriere$^{c}$, J.~Leon$^{a}$, J.~MacMullin$^{n,i}$, R.D.~Martin $^{j}$, S.J.~Meijer$^{n,i}$, S.~Mertens$^{b}$, J.L.~Orrell$^{c}$, C.~O'Shaughnessy$^{n,i}$, A.W.P.~Poon$^{b}$, D.C.~Radford$^{e}$, J.~Rager$^{n,i}$, K.~Rielage$^{o}$, R.G.H.~Robertson$^{a}$, E.~Romero-Romero$^{l,e}$, B.~Shanks$^{n,i}$, M.~Shirchenko$^{g}$, N.~Snyder$^{j}$, A.M.~Suriano$^{k}$, D.~Tedeschi$^{d}$, J.E.~Trimble$^{n,i}$, R.L.~Varner$^{e}$, S. Vasilyev$^{g}$, K.~Vetter$^{r,b}$, K.~Vorren$^{n,i}$, B.R.~White$^{e}$, J.F.~Wilkerson$^{n,i,e}$, C. Wiseman$^{d}$, W.~Xu$^{o}$, E.~Yakushev$^{g}$, C.-H.~Yu$^{e}$, V.~Yumatov$^{f}$, I.~Zhitnikov$^{g}$}
{address={$^{a}$ \uw   \\
          $^{b}$ \lbnl  \\
          $^{c}$  \pnnl \\
          $^{d}$  \usc \\
          $^{e}$  \ornl \\
          $^{f}$  \ITEP \\
          $^{g}$ \JINR  \\
          $^{h}$  \duke \\
          $^{i}$  \tunl\\
          $^{j}$  \usd \\
          $^{k}$  \sdsmt \\
          $^{l}$  \ut  \\
          $^{m}$  \ou  \\
          $^{n}$ \unc \\
          $^{o}$  \lanl \\
          $^{p}$  \blhill \\
          $^{q}$  \ttu \\
          $^{r}$  Alternate address: Department of Nuclear Engineering, University of California, Berkeley, CA, USA
          }}	

}

%% file: LMFEBGTable.tex
\begin{tabular}{@{} ccccc @{}} 
      \hline
      \hline
      \multirow{2}*{Material}	& Assay	& \multirow{2}*{Isotope}& purity	& MJD BG \\
      				& Method&			& [pg/g] & [c/ROI/t/y]	\\
      \hline
      \hline
      \multirow{2}*{Fused Silica} 	& \multirow{2}*{ICP-MS}	& $^{238}$U & 284	& 0.0616 \\
      				   	&			& $^{232}$Th& 101	& 0.0259 \\
      \hline
      \multirow{2}*{aGe}		& \multirow{2}*{ICP-MS}	& $^{238}$U & 5000	& 0.0001 \\
      					&			& $^{232}$Th& 5000	& 0.0001 \\
      \hline
      \multirow{2}*{Au}			& \multirow{2}*{ICP-MS}	& $^{238}$U & 2000	& 0.0015 \\
      					&			& $^{232}$Th& 47000	& 0.0421 \\
      \hline
      \multirow{2}*{Ti}			& \multirow{2}*{ICP-MS}	& $^{238}$U & $<$100	& $\sim$0 \\
      					&			& $^{232}$Th& $<$400	& $\sim$0 \\
      \hline
      \multirow{2}*{FET die}		& \multirow{2}*{ICP-MS}	& $^{238}$U & $<$141	& $<$0.0006 \\
      					&			& $^{232}$Th& $<$2000	& $<$0.0107 \\
      \hline
      \multirow{2}*{Al}			& \multirow{2}*{ICP-MS}	& $^{238}$U & 91000	& 0.0004 \\
      					&			& $^{232}$Th& 9.0	& $\sim$0 \\
      \hline
      \multirow{2}*{Ag epoxy}		& \multirow{2}*{$\gamma$-counting}	& $^{238}$U & $<$10000	& $<$0.0082 \\
      					&			& $^{232}$Th& $<$70000	& $<$0.0685 \\
      \hline
      \multirow{2}*{EFCu Spring Clip}	& \multirow{2}*{ICP-MS}	& $^{238}$U & $<$0.015	&  0.0005\\
      					&			& $^{232}$Th& $<$0.014	&  0.0003\\
      \hline
      \hline
\end{tabular}

%% file: PlugBGTable.tex
\begin{tabular}{@{} cccccc @{}} 
  \hline
  \hline
  \multirow{2}*{Material} 	& Assay  & Mass	[g per	& \multirow{2}*{Isotope}  & Activity		& MJD BG \\
  				& Method & conn. pair] 	&	 		  & [$\mu$Bq/kg] 	& [c/ROI/t/y]	\\
  \hline
  \hline
  \multirow{2}*{Pins (w/BeCu)} 	& \multirow{2}*{ICP-MS}	& \multirow{2}*{0.112}	& $^{238}$U & $795000 \pm 12000$ & $8.8 \pm  0.1$ \\
  			&	&			& $^{232}$Th		& $41000 \pm 1000$	& $2.3 \pm 0.1$ \\
  \hline
  \multirow{2}*{Pins (no BeCu)}	& \multirow{2}*{ICP-MS}	& \multirow{2}*{0.112}	& $^{238}$U & $4600 \pm 1500$ 	& $0.05 \pm 0.02$ \\
  				&			&			& $^{232}$Th& $5800 \pm 100$  	& $0.32 \pm 0.01$ \\
  \hline
  \multirow{2}*{Vespel SP-1}	& \multirow{2}*{NAA}	& \multirow{2}*{0.95}	& $^{238}$U & $<$1000		& $<$0.20 \\
	  			&			&			& $^{232}$Th& $<$12		& $<$0.01 \\
  \hline
  \multirow{2}*{Solder}		& \multirow{2}*{GDMS}	& \multirow{2}*{0.04}	& $^{238}$U & $5600 \pm 1000$	& $0.02 \pm 0.004$ \\
  				&			&			& $^{232}$Th& $<$12		& $<$0.0002 \\
  \hline
  \multirow{2}*{Solder flux}	& \multirow{2}*{GDMS}	& \multirow{2}*{0.04}	& $^{238}$U & $1200 \pm 200$ 	& $0.005 \pm 0.001$ \\
 	 			&			&			& $^{232}$Th& $<$400		& $<$0.007 \\
  \hline
  \multirow{2}*{FEP (shrink tube)}& \multirow{2}*{NAA}	&\multirow{2}*{$<$0.001}&$^{238}$U & $<$1250 		& $<$0.0001 \\
 	 			&			&			& $^{232}$Th& $<$138		& $<$0.0001 \\
  \hline
  \multirow{2}*{Total}		&			& \multirow{2}*{1.14}	& $^{238}$U & $<$1520		& $<$0.28 \\
  				&			&			& $^{232}$Th& $<$593		& $<$0.34 \\
  \hline
  \hline
\end{tabular}